\documentstyle[epsfig,aps,prd]{revtex}
\begin{document}
\bibliographystyle{plain}
\thispagestyle{empty}
\vspace{-5mm}
\bigskip
%\bigskip
\begin{center}
\large{\bf  Renormalization Group Evolution and 
Infra-red Fixed Points in Minimal Supersymmetric Standard Model
with Baryon and Lepton Number Violation}

\vskip 1.5cm

B. Ananthanarayan\\
Centre for Theoretical Studies,\\
Indian Institute of Science,\\ 
Bangalore 560 012, India 

\vskip 1.0cm

P. N. Pandita\\
Department of Physics,\\
North-Eastern Hill University,\\
Shillong 793 022, India

\end{center}

\vskip 1.5cm

\begin{abstract}
We study the renormalization group evolution and infra-red stable fixed 
points of the Yukawa couplings and the soft 
supersymmetry breaking trilinear couplings of the minimal
supersymmetric standard model with baryon and lepton number (and R-parity)
violation involving the heaviest generations.  
We show analytically  that in the Yukawa 
sector there is only
one infra-red stable fixed point. This corresponds
to non-trivial fixed point for the top-, bottom-quark Yukawa couplings and the
$B$ violating coupling $\lambda_{233}''$, and a trivial one for
the $\tau$-Yukawa coupling.  All other possible
fixed points are either unphysical or unstable in the infra-red region.  
We then carry out an analysis of the renormalization group
equations for the soft supersymmetry breaking trilinear 
couplings, and determine the corresponding fixed points for these
couplings. We also study the quasi-fixed point behaviour, both 
algebraically and with numerical solutions of the evolution equations, 
of the third generation Yukawa couplings and the baryon number violating
coupling, and those of the soft supersymmetry breaking trilinear couplings.
From the analysis of the fixed point behaviour, we obtain upper and
lower bounds on the baryon number violating coupling $\lambda_{233}''$,
as well as on the supersymmetry breaking trilinear couplings.
\end{abstract}

\bigskip

PACS number(s):  11.10.Hi, 11.30.Fs, 12.60.Jv

%{\it Keywords:  Supersymmetry, R-parity violation, Infra-red fixed points}

\newpage

\section{Introduction}

The Standard Model (SM) is a tremendous success in describing
the strong and electroweak interactions based on an underlying gauge
principle.  However, one of the main weaknesses of the SM is that
the masses of the matter particles, 
the quarks and leptons,  are
free parameters of the theory.  This weakness persists in most extensions
of the SM, including the supersymmetric extensions of the Standard Model.  
The fermions mass problem in the SM and its extensions arises from the 
presence of many unknown dimensionless Yukawa couplings.  
On the other hand the
minimal supersymmetric version of the SM leads to a successful prediction
for the ratio of the gauge couplings with a gauge unificiation scale 
$M_G \simeq 10^{16}$ GeV.  This has led to the idea that there may
be a stage of unification beyond the SM.  If so, then it becomes important
to perform the radiative corrections in determining all the dimension
$\leq 4$ terms in the lagrangian.  This can be achieved by using the 
renormalization group equations in finding the values of parameters
at the low scale, given their value at a high scale.
Thus, considerable attention has recently
been focussed on the renormalization group evolution~\cite{schrempp1}
of the various dimensionless Yukawa couplings in the SM and its minimal
supersymmetric extension, the Minimal Supersymmetric Standard Model (MSSM).
Using the renormalization group evolution, one may attempt to relate the
Yukawa couplings to the gauge couplings via the Pendleton-Ross infra-red
stable fixed point (IRSFP) for the top-quark Yukawa coupling~\cite{pendross},
or via the quasi-fixed point behaviour~\cite{hill}.  The predictive power
of the SM and its supersymmetric extensions may, thus, be enhanced if
the renormalization group (RG) running of the parameters is dominated by
infra-red stable fixed  points (IRSFPs).  
Typically, these fixed points are for ratios like Yukawa couplings
to the gauge coupling, or in the context of supersymmetric models, the 
supersymmetry breaking trilinear $A$-parameter to the gaugino mass, etc.
These ratios do not attain their fixed point values at the weak scale, the
range between the GUT (or Planck) scale and the weak scale being too small
for the ratios to closely approach the fixed point.  Nevertheless, the
couplings may be determined by quasi-fixed point behaviour~\cite{hill}
where the value of the Yukawa coupling at the weak scale is independent of
its value at the GUT scale, provided the Yukawa coupling at the GUT scale
is large.  For the fixed point or the quasi-fixed point scenarios to
be successful, it is necessary that these fixed points  are
stable~\cite{allanach,abel,jack}.  

Since supersymmetry~\cite{nilles} requires the introduction of superpartners of
all known particles in the SM (in addition to the introduction of at least
two Higgs doublets), which transform in an identical manner under the gauge
group, there are additional Yukawa couplings in supersymmetric models
which violate~\cite{weinberg} baryon number~($B$) or lepton number~($L$).
In MSSM,  a discrete symmetry called R-parity ($R_p$) is invoked to
eliminate these $B$ and $L$ violating Yukawa couplings~\cite{farrar}.
However, the assumption of $R_p$ conservation at the level of the MSSM
appears to be {\it ad hoc}, since it is not required for the internal
consistency of the model.  Therefore, the study of the renormalization
group evolution of the dimensionless Yukawa couplings in  the MSSM,
including $B$ and $L$ (and $R_p$) violation, deserves serious consideration.

Recently considerable attention has been focussed on the study of the
RG evolution of the top- and bottom-quark Yukawa couplings,  
and the $R_p$ violating Yukawa couplings,  of the MSSM.  
This includes the study of quasi-fixed point behaviour~\cite{barger} as
well as the true infra-red fixed points of different Yukawa couplings,
and the analysis of their stability~\cite{ap1}.  This has led to certain
insights and constraints on the fixed point behaviour of some of the $R_p$
violating couplings, involving higher generation indices.  
As pointed out earlier, for the 
fixed point or the quasi-fixed point behaviour to be successful, it is 
necessary that the fixed points are stable~\cite{ap1,allanach}.  
It has been shown that in MSSM with only top- and bottom-quark
Yukawa couplings, and with the highest generation $B$ or $L$ violation
taken into account, only the $B$-violating
coupling $\lambda''_{233}$, together with top- and bottom-quark
Yukawa couplings,
approaches a non-trivial IRSFP, whereas all other non-trivial fixed
points are either unphysical or unstable in the infra-red region~\cite{ap1}.
This conclusion remains 
unchanged even if one extends the MSSM by including a 
Higgs singlet superfield, to  the so-called non-minimal supersymmetric
standard model~\cite{paulraj}.

In this paper we carry out a detailed study of the 
renormalization group evolution of the Yukawa couplings of MSSM, 
including the $B$ and $L$ violating couplings.  
We shall include all the third generation Yukawa couplings~\cite{schrempp2}, 
as well as the highest generation $B$ and $L$ violating
couplings in our study, and analyze the situation where 
all of them could simultaneously approach infra-red fixed points.  
We shall investigate both the true infra-red
fixed points, as well as quasi-fixed points of these couplings.
In particular, we shall carry out a detailed stability analysis of the
infra-red fixed points of these couplings.  
Furthermore, corresponding to the $B$ and $L$~(and $R_p$)  violating
Yukawa couplings of the MSSM, there are soft supersymmetry
breaking trilinear couplings (the $A$ parameters)
whose renormalization group evolution
and infra-red fixed point structure has not been studied so far.  
We shall, therefore,  also study the renormalization
group evolution of these soft supersymmetry breaking $A$ parameters, 
including those corresponding to the third generation Yukawa couplings, 
and obtain the simultaneous infra-red fixed points for them.

The plan of the paper is as follows.  In Sec. II we describe the model
and write down the renormalization group equations of interest to us,
which we have rederived since we require the complete set of equations 
involving baryon and lepton number violation.
We then carry out a detailed analytical study of the 
true infra-red fixed points of the Yukawa couplings in 
full generality. Within the context of grand unified theories, one is
led to the situation where $B$ and $L$ violating Yukawa couplings may be 
related at the GUT scale, and one may no longer be able to set 
one or the other arbitarily to zero. We, therefore, initially 
include both baryon and lepton number violation in our RG equations.
The fixed point analysis of such a system of RG equations
leads to the crucial result that
the  only stable fixed point is the one with 
simultaneous non-trivial 
fixed point values for the top- and bottom-quark Yukawa couplings and 
the $B$-violating coupling $\lambda_{233}''$, and a trivial one for the
$\tau$-Yukawa coupling. Thus, non-trivial 
simultaneous fixed points for the $B$ and $L$ violating Yukawa couplings
are ruled out by our analysis.
We then study
the fixed points of the corresponding soft supersymmetry breaking 
trilinear couplings of this model.  In Sec. III
we algebraically study the simultaneous 
quasi-fixed points of all the third 
generation Yukawa couplings of the minimal 
supersymmetric standard model with $B$ violation, 
as well as those of the corresponding soft supersymmetry 
breaking trilinear couplings. Since the quasi-fixed point
limit is formally defined as the Landau pole of the 
Yukawa   coupling at the GUT scale, it provides an upper bound
on the corresponding Yukawa coupling. 
In Sec. IV we present the numerical results 
for the renormalization group evolution and the quasi-fixed points
for the minimal supersymmetric standard model with $B$ 
violation.  In Sec. V we summarize  our results and
present the conclusions.

\section{Renormalization Group Equations and Infra-red Fixed Points}

In this section we study the true infra-red fixed points of the 
Yukawa couplings and the $A$ parameters of the MSSM with $B$ and $L$
violation.  We begin by recalling the basic features of the model.
The superpotential of the MSSM is written as  
\begin{equation}\label{mssmsuperpotential}
W= (h_U)_{ab} Q^a_L \overline{U}^b_R H_2 
+ (h_D)_{ab} Q^a_L \overline{D}^b_R H_1 
+ (h_E)_{ab} L^a_L \overline{E}^b_R H_1 + \mu H_1 H_2,
 \end{equation}
where $L,\, Q, \, \overline{E},\, \overline{D},\, \overline{U}$
denote the lepton and quark doublets, and anti-lepon singlet, d-type 
anti-quark singlet and u-type anti-quark singlet,  respectively.
In Eq.~(\ref{mssmsuperpotential}), $(h_U)_{ab}$, $(h_D)_{ab}$
and $(h_E)_{ab}$ are the Yukawa coupling matrices, with $a,\, b,\, c$
as the generation indices.
Gauge invariance, supersymmetry and renormalizability allow the addition of
the following $L$ and $B$ violating terms to the 
MSSM superpotential (\ref{mssmsuperpotential}):
\begin{eqnarray}
& \displaystyle W_L= {1\over 2}\lambda_{abc} L^a_L L^b_L \overline{E}^c_R
+ \lambda'_{abc} L^a_L Q^b_L\overline{D}^c_R
+ \mu_i L_i H_2, & \label{Lviolating} \\
& \displaystyle W_B={1\over 2}\lambda''_{abc} \overline{D}^a_R
\overline{D}^b_R \overline{U}^c_R,  & \label{Bviolating} 
\end{eqnarray}
respectively.  The Yukawa couplings
$\lambda_{abc}$ and $\lambda''_{abc}$ are antisymmetric in their
first two indices due to $SU(2)_L$ and $SU(3)_C$ group structures,
respectively.  Corresponding to the terms in the superpotentials
(\ref{mssmsuperpotential}), (\ref{Lviolating}) and (\ref{Bviolating}),
there are the soft supersymmetry breaking trilinear terms 
which can be written as
\begin{eqnarray}
& \displaystyle -V_{\rm{soft}}=  
\left[(A_U)_{ab}(h_U)_{ab} \tilde{Q}^a_L \tilde{\overline{U}}^b_R H_2 
+ (A_D)_{ab}(h_D)_{ab} \tilde{Q}^a_L \tilde{\overline{D}}^b_R H_1 
+ (A_E)_{ab}(h_E)_{ab} \tilde{L}^a_L \tilde{\overline{E}}^b_R H_1\right] 
& \nonumber \\
& \displaystyle 
+ \left[{1\over 2}(A_\lambda)_{abc}\lambda_{abc} \tilde{L}^a_L \tilde{L}^b_L 
\tilde{\overline{E}}^c_R
+ (A_{\lambda'})_{abc}\lambda'_{abc} \tilde{L}^a_L \tilde{Q}^b_L
\tilde{\overline{D}}^c_R \right] + 
 \left[ {1\over 2}(A_{\lambda''})_{abc}\lambda''_{abc} \tilde{\overline{D}}^a_R
\tilde{\overline{D}}^b_R \tilde{\overline{U}}^c_R \right], & \label{soft}
\end{eqnarray}
where a tilde denotes the scalar component of the chiral superfield, and
the notation for the scalar component of the Higgs superfield is the same
as that of the corresponding superfield. In addition there are
soft supersymmetry breaking gaugino
mass terms with the masses $M_i$, with $i = 1, 2, 3$
corresponding to the gauge groups $U(1)_Y, \, SU(2)_L, \,$ and 
$SU(3)_C$, respectively.

The third generation Yukawa couplings are the dominant couplings in the
superpotential (\ref{mssmsuperpotential}), so it is natural
to retain only the elements $(h_U)_{33}\equiv h_t$,  
$(h_D)_{33} \equiv h_b$, $(h_L)_{33} \equiv h_\tau$ 
in each of the Yukawa couplings matrices $h_U,\, h_D,
\, h_L$, setting all other elements equal to zero.  Furthermore, there are
36 independent $L$ violating trilinear couplings $\lambda_{abc}$ and
$\lambda'_{abc}$ in (\ref{Lviolating}).  Similarly, there are
9 independent $B$ violating couplings $\lambda''_{abc}$ in 
the baryon number violating superpotential (\ref{Bviolating}).
Thus, we would have to consider 39 coupled nonlinear evolution equations
for the $L$ violating case and 12 coupled nonlinear equations for the $B$
violating case, respectively.  It is clear that
there is a need for a radical
simplification of these equations before we can think of studying the
evolution of the Yukawa couplings in the MSSM 
with  $B$ and $L$ violation.

In order to render the Yukawa couping evolution equations 
tractable, we, therefore,  need
to make certain plausible assumptions.  Motivated by the generational
hierarchy of the conventional Higgs couplings, we shall assume that an
analogous hierarchy amongst the different generations of $B$ and $L$
violating couplings exists.  Thus, we shall retain only the 
couplings $\lambda_{233}, \, \lambda'_{333},\,  \lambda''_{233}$,
and neglect the rest.  We note that $B$ and $L$
violating couplings
to higher generations evolve more strongly because of larger Higgs
couplings in their evolution equations, and hence could take larger
values than the corresponding couplings to the lighter generations.
We also note that the experimental upper limits are stronger for the
$B$ and $L$ violating couplings with lower indices~\cite{barbier}.

With these assumptions the one-loop renormalization group 
equations~\cite{rges} for the Yukawa couplings,  
and the $B$ and $L$ violating couplings,
following from the various terms in the superpotentials
(\ref{mssmsuperpotential}),
(\ref{Lviolating}) and (\ref{Bviolating}), respectively,  
can be written as
\begin{eqnarray}
%%%%%%%%%%%%%%%%%%%%%%%%%%%%%%%%%%%%%%%%%%
& \displaystyle {dh_t\over d\ln \mu}={h_t\over 16 \pi^2}
\left(6 h_t^2 + h_b^2 + \lambda'^2_{333} + 2 \lambda''^2_{233}-
{16\over 3} g_3^2 - 3 g_2^2 - {13\over 15}g_1^2 \right),  & \label{rge1}\\
%%%%%%%%%%%%%%%%%%%%%%%%%%%%%%%%%%%%%%%%%%
& \displaystyle {dh_b\over d\ln \mu}={h_b\over 16 \pi^2}
\left( h_t^2 + 6 h_b^2 + h_\tau^2 + 6\lambda'^2_{333} + 2 \lambda''^2_{233}-
{16\over 3} g_3^2 - 3 g_2^2 - {7\over 15}g_1^2 \right),  & \label{rge2}\\
%%%%%%%%%%%%%%%%%%%%%%%%%%%%%%%%%%%%%%%%%%
& \displaystyle {dh_\tau\over d\ln \mu}={h_\tau\over 16 \pi^2}
\left( 3 h_b^2 + 4 h_\tau^2 + 4\lambda_{233}^2 + 3 \lambda'^2_{333}
 - 3 g_2^2 - {9\over 15}g_1^2 \right),  & \label{rge3}\\
%%%%%%%%%%%%%%%%%%%%%%%%%%%%%%%%%%%%%%%%%%
& \displaystyle {d\lambda_{233}\over d\ln \mu}={\lambda_{233}\over 16 \pi^2}
\left(  4 h_\tau^2 + 4\lambda_{233}^2 + 3 \lambda'^2_{333}
 - 3 g_2^2 - {9\over 15}g_1^2 \right),  & \label{rge4}\\
%%%%%%%%%%%%%%%%%%%%%%%%%%%%%%%%%%%%%%%%%%
%%%%%%%%%%%%%%%%%%%%%%%%%%%%%%%%%%%%%%%%%%
& \displaystyle {d\lambda'_{333}\over d\ln \mu}={\lambda'_{333}\over 16 \pi^2}
\left( h_t^2 + 6 h_b^2 + h_\tau^2 + \lambda_{233}^2+
 6\lambda'^2_{333} + 2 \lambda''^2_{233}-
{16\over 3} g_3^2 - 3 g_2^2 - {7\over 15}g_1^2 \right),  & \label{rge5}\\
%%%%%%%%%%%%%%%%%%%%%%%%%%%%%%%%%%%%%%%%%%
& \displaystyle {d\lambda''_{233}\over d\ln \mu}={\lambda''_{233}\over 16 \pi^2}
\left(2 h_t^2 + 2 h_b^2 + 2 \lambda'^2_{333} + 6 \lambda''^2_{233}-
8 g_3^2  - {4\over 15}g_1^2 \right),  & \label{rge6}
%%%%%%%%%%%%%%%%%%%%%%%%%%%%%%%%%%%%%%%%%%
\end{eqnarray}
where $g_1,\, g_2, \, g_3$ are the gauge couplings of $U(1)_Y$
(in the GUT normalization), $SU(2)_L$
and $SU(3)_C$ gauge groups, respectively,
and $\mu$ is the scale parameter.  The
evolution equations for the gauge couplings are not affected by the 
presence of $B$ and $L$
violating couplings at the one-loop level, and can be
written as
\begin{equation} \label{gaugerge}
16\pi^2 {dg_i\over d\ln\mu}=b_i g_i^3, \, \, \, i=1, \, 2, \, 3,
\end{equation}
with 
\begin{equation}
b_1 = 33/5, \, b_2 = 1, \, b_3 = -3.
\end{equation}
The corresponding one-loop renormalization group equations for the gaugino
masses $M_i,\, i=1, \, 2, \, 3$ can be written as
\begin{equation} \label{gauginorge}
16 \pi^2 {d M_i\over d\ln\mu}=  2 g_i^2 b_i M_i.
\end{equation}
We now come to the evolution equations for the soft supersymmetry
breaking trilinear parameters in potential 
(\ref{soft}).  The one-loop RGEs 
for these parameters can be deduced from the general expressions in
ref.~\cite{rges}.  In this paper
we shall assume the same kind of generational
hierarchy for these trilinear parameters as was assumed for the
corresponding Yukawa couplings.  Thus,  we shall consider only the 
highest generation trilinear coulings 
$(A_U)_{33}\equiv A_t$,  $(A_D)_{33}\equiv A_b$, $(A_L)_{33}\equiv A_\tau$,
$(A_\lambda)_{233}\equiv A_\lambda$, 
$(A_{\lambda'})_{333}\equiv A_{\lambda'}$,
$(A_{\lambda''})_{233}\equiv A_{\lambda''}$, setting all other elements
equal to zero.  With this assumption the RGEs for the soft supersymmetry 
breaking trilinear parameters can be written as
\begin{small}
\begin{eqnarray}
%%%%%%%%%%%%%%%%%%%%%%%%%%%%%%%%%%%%%%%%%%
& \displaystyle {dA_t\over d\ln \mu}={1\over 8 \pi^2}
\left(6 A_t h_t^2 + A_b h_b^2 + A_{\lambda'}
\lambda'^2_{333} + 2 A_{\lambda''} \lambda''^2_{233}-
{16\over 3}M_3 g_3^2 - 3 M_2 g_2^2 - {13\over 15} M_1 g_1^2 \right),  & 
\label{rge7}\\
%%%%%%%%%%%%%%%%%%%%%%%%%%%%%%%%%%%%%%%%%%
& \displaystyle {dA_b\over d\ln \mu}={1\over 8\pi^2}
\left(A_t h_t^2 + 6 A_b h_b^2 + A_\tau h_\tau^2 + 
6 A_{\lambda'}\lambda'^2_{333} + 2 A_{\lambda''}\lambda''^2_{233}-
{16\over 3}M_3 g_3^2 - 3 M_2 g_2^2 - {7\over 15}M_1 g_1^2 \right),  &
 \label{rge8}\\
%%%%%%%%%%%%%%%%%%%%%%%%%%%%%%%%%%%%%%%%%%
& \displaystyle {dA_\tau\over d\ln \mu}={1\over 8 \pi^2}
\left( 3 A_b h_b^2 + 4 A_\tau h_\tau^2 + 4A_\lambda
\lambda_{233}^2 + 3 A_{\lambda'}\lambda'^2_{333}
 - 3 M_2 g_2^2 - {9\over 15}M_1 g_1^2 \right),  & \label{rge9}\\
%%%%%%%%%%%%%%%%%%%%%%%%%%%%%%%%%%%%%%%%%%
& \displaystyle {d A_\lambda\over d\ln \mu}={1\over 8 \pi^2}
\left(  4 A_\tau h_\tau^2 + 4A_\lambda\lambda_{233}^2 + 3 A_{\lambda'}
\lambda'^2_{333}
 - 3 M_2 g_2^2 - {9\over 15}M_1 g_1^2 \right),  & \label{rge10}\\
%%%%%%%%%%%%%%%%%%%%%%%%%%%%%%%%%%%%%%%%%%
%%%%%%%%%%%%%%%%%%%%%%%%%%%%%%%%%%%%%%%%%%
& \displaystyle {dA_{\lambda'}\over d\ln \mu}={1\over 8 \pi^2}
\left( A_t h_t^2 + 6 A_b h_b^2 + A_\tau h_\tau^2 + A_\lambda\lambda_{233}^2+
 6 A_{\lambda'}\lambda'^2_{333} + 2 A_{\lambda''}\lambda''^2_{233}-
{16\over 3} M_3 g_3^2 - 3 M_2 g_2^2 - {7\over 15}M_1 g_1^2 \right),  & 
\label{rge11}\\
%%%%%%%%%%%%%%%%%%%%%%%%%%%%%%%%%%%%%%%%%%
& \displaystyle {dA_{\lambda''}\over d\ln \mu}={1\over 8 \pi^2}
\left(2 A_t h_t^2 + 2 A_b h_b^2 + 2 A_{\lambda'}
 \lambda'^2_{333} + 6 A_{\lambda''}\lambda''^2_{233}-
8 M_3 g_3^2  - {4\over 15}M_1 g_1^2 \right).  & \label{rge12}
%%%%%%%%%%%%%%%%%%%%%%%%%%%%%%%%%%%%%%%%%%
\end{eqnarray}
\end{small}
Given the evolution equations (\ref{rge1}) - (\ref{rge6}) for the
Yukawa couplings and the  evolution 
equations (\ref{rge7}) - (\ref{rge12}) for the
$A$ parameters, we are now ready to study the RG evolution
and infra-red fixed 
points of these parameters of MSSM with $B$ and $L$ violation.

\subsection{Infra-red Fixed Points for Yukawa Couplings}
We first consider  IRFPs for  the Yukawa couplings.
With the definitions
\begin{equation}\label{redefinitions}
R_t={h_t^2\over g_3^2}, \,\,\, R_b={h_b^2\over g_3^2}, \,\,\,
R_\tau={h_\tau^2\over g_3^2},  \,\,\, 
R={\lambda^2_{233}\over g_3^2}, \, \, \,
R'={\lambda'^2_{333}\over g_3^2}, \, \, \,
R''={\lambda''^2_{233}\over g_3^2}, \, \, \,
\end{equation}
and retaining only the $SU(3)_C$ gauge coupling constant, we can
rewrite the renormalization group equations for the Yukawa couplings
as ($\tilde{\alpha}_3 =g_s^2/(16\pi^2)$)
\begin{eqnarray}
& \displaystyle {dR_t\over dt}=\tilde{\alpha_3} R_t
\left[\left({16\over 3}+b_3\right)-
6 R_t - R_b - R'- 2 R''\right],  & \label{rtequation}
 \\
& \displaystyle {dR_b\over dt}=\tilde{\alpha_3} R_b
\left[\left({16\over 3}+b_3\right)
- R_t -6 R_b -R_\tau- 6 R'- 2 R''\right],  & \label{rbequation}
 \\
& \displaystyle {dR_\tau\over dt}=\tilde{\alpha_3} R_\tau
\left[b_3 -3 R_b -4 R_\tau -4 R - 3 R'\right],  & \label{rtauequation}
\\
& \displaystyle {dR\over dt}=\tilde{\alpha_3} R
\left[b_3 -4 R_\tau -4 R -3 R' \right], & \label{requation} \\
& \displaystyle {dR'\over dt}=\tilde{\alpha_3} R'
\left[\left({16\over 3}+b_3\right)
- R_t -6 R_b -R_\tau -R - 6 R'- 2 R''\right], & \label{rprimeequation} \\
& \displaystyle {dR''\over dt}=\tilde{\alpha_3} R''
\left[\left(8+b_3\right)
-2 R_t -2 R_b - 2 R'- 6 R''\right], & \label{rdoubleprimeequation}
\end{eqnarray}
where $b_3=-3$, the beta function for $g_3$ in the MSSM, and
$t=-\ln\, \mu^2$.

The renormalization group evolution~\cite{barger} and the infra-red fixed
points~\cite{ap1} of the set of equations
(\ref{rtequation}) - (\ref{rdoubleprimeequation}) has been studied
in the limit of ignoring the $\tau$-Yukawa coupling $h_\tau$, and
by considering either the baryon number violating Yukawa coupling
$\lambda''_{233}$, or the lepton number violating
Yukawa couplings $\lambda_{233}$
and $\lambda'_{333}$.  In the analysis that follows, we shall consider
the evolution equations for $h_t$ and $h_b$
together with the evolution equation
for $h_{\tau}$.  Furthermore, we shall also entertain the
possibility of simultaneous presence of $B$ and $L$ violating couplings in the
renormalization group equations
(\ref{rtequation}) - (\ref{rdoubleprimeequation}).  
We do this in order to investigate as to whether such a system of equations
does have acceptable infra-red fixed points.  Ordering the ratios
as $R_i=(R'',R',R,R_\tau,R_b,R_t)$, we rewrite the RGEs 
(\ref{rtequation}) - (\ref{rdoubleprimeequation}) in the form
\begin{equation}\label{componentequation}
{d R_i\over dt}=\tilde{\alpha}_3 R_i
\left[(r_i+b_3)-\sum_j S_{ij} R_j\right],
\end{equation}
where $r_i=\sum_R 2 \, C_R$,  $C_R$ is the QCD Casimir
for the various fields ($C_Q=C_{\overline{U}}=C_{\overline{D}}=4/3$), 
the sum is over the representation of
the three fields associated with the trilinear coupling that enters
$R_i$, and $S$ is a matrix whose value is fully specified by
the wavefunction anomalous dimensions. 
A fixed point is,  then,  reached
when the right hand side of Eq.~(\ref{componentequation}) is
0 for all $i$.  If we were to write the fixed point solutions as
$R_i^*$, then there are two fixed point values for each coupling:
$R_i^*=0$, or
\begin{equation}
\left[\left(r_i+b_3\right) -\sum_j S_{ij} R_j^* \right]=0.
\end{equation}
It follows that the non-trivial fixed point solution is
\begin{equation}\label{ristarequation}
R_i^*=\sum_j (S^{-1})_{ij} (r_j+b_3).
\end{equation}
The anomalous dimension matrix S that enters Eq.~(\ref{ristarequation}),  
which we denote by $S_{BL}$, is readily seen to be
\begin{equation}\label{sbl}
S_{BL}=\left[
\begin{array}{ c c c c c c}
6 & 2 & 0 & 0 & 2 & 2 \\
2 & 6 & 1 & 1 & 6 & 1 \\
0 & 3 & 4 & 4 & 0 & 0 \\
0 & 3 & 4 & 4 & 3 & 0 \\
2 & 6 & 0 & 1 & 6 & 1 \\
2 & 1 & 0 & 0 & 1 & 6 \\
\end{array}
\right].
\end{equation}
Inverting the matrix  (\ref{sbl}) and substituting in
Eq.(\ref{ristarequation}), we get
the following fixed point solution:
\begin{eqnarray}
& \displaystyle R''^*={611\over 876}, \, \, \, R'^*={22/73}, \, \, \, 
R^*=0,  & \nonumber \\
& \displaystyle R^*_\tau=-{285\over 292}, \, \, \, R^*_b=0, \, \, \,
R_t^*={31\over 292}. &
\end{eqnarray}
We note that $R^*_\tau<0$, and, therefore this fixed point solution is 
unacceptable.  We,  therefore,  conclude  that a simultaneous fixed
point for the $B$ and $L$  violating couplings $\lambda''_{233}$,
and $\lambda'_{333}, \, \lambda_{333}$,  and the
Yukawa couplings $h_\tau, \, h_b, \, h_t$ does
not exist.

We next consider the possibility of one of the $L$ violating couplings
attaining a zero fixed point value, with all others having non-trivial
fixed point values.  We first consider the case with $R^*=0$.
The corresponding anomalous dimension matrix we need to consider,
denoted by $\tilde{S}_{BL}$, is
\begin{equation}
\tilde{S}_{BL}=\left[
\begin{array}{c c c c c}
6 & 2 & 0 & 2 & 2 \\
2 & 6 & 1 & 6 & 1 \\
0 & 3 & 4 & 3 & 0 \\
2 & 6 & 1 & 6 & 1 \\
2 & 1 & 0 & 1 & 6 \\
\end{array}
\right],
\end{equation}
which is singular.  Hence there are no fixed points in this case.

We next consider the possibility of having $R'^*=0$, with 
all other Yukawa couplings having non-trivial fixed points.  The anomalous
dimension matrix in this case is denoted by $\tilde{\tilde{S}}_{BL}$,
and is given by
\begin{equation}
\tilde{\tilde{S}}_{BL}=\left[
\begin{array}{c c c c c}
6 & 0 & 0 & 2 & 2 \\
0 & 4 & 4 & 0 & 0 \\
0 & 4 & 4 & 3 & 0 \\
2 & 0 & 1 & 6 & 1 \\
2 & 0 & 0 & 1 & 6 \\
\end{array}
\right],
\end{equation}
leading to the fixed point values
\begin{eqnarray}
& \displaystyle R''^*={19\over 24}, \,  \, \, R^*=-{11\over 8}, & \nonumber \\
& \displaystyle R^*_\tau={5\over 8}, \, \, \, R^*_b=0, \, \, \,
R_t^*={1\over 8}, &
\end{eqnarray}
which is physically unacceptable.  We conclude that $B$ and $L$
violating couplings
of the highest generation cannot simultaneously approach a 
non-trivial fixed point in
the MSSM.  This is, perhaps, one of the most important conclusions  that
we draw from the analysis of the renormalization group equations in 
this paper.

It is now natural to consider the possibility of having either 
$B$ or  $L$
violation, but not both simultaneously,  involving the 
Yukawa couplings with  highest generation
indices in the renormalization group evolution 
in the MSSM.

\subsubsection{Fixed Points with $B$ violation}

In this section we consider the possibility of having simultaneous 
fixed pints for
the Yukawa couplings $h_\tau, \, h_b, \, h_t$, and the $B$
violating coupling
$\lambda''_{233}$.  Ordering the ratios of the Yukawa couplings to
the gauge couplings as $R_i=(R_\tau, R'', R_b, R_t)$, the anomalous
dimension matrix, denoted by $S_B$, can be written as
\begin{equation}
S_B=\left[
\begin{array}{c c c c}
4 & 0 & 3 & 0 \\
0 & 6 & 2 & 2 \\
1 & 2 & 6 & 1 \\
0 & 2 & 1 & 6 \\
\end{array}
\right],
\end{equation}
leading to the fixed point values
\begin{eqnarray}
& \displaystyle R_\tau^*=-{285\over 292}, \, \, \, R''^*={611\over 876},
& \nonumber \\
& \displaystyle R_b^*={22\over 73}, \, \, \, R_t^*={31\over 292}, &
\end{eqnarray}
which must be rejected as being unphysical.  We are, therefore, led to
the consideration of a fixed point with one of the Yukawa couplings 
approaching a zero fixed point value,
with all others attaining a non-trivial fixed point.  
We try the fixed point with $R^*_\tau=0$, with all others
obtaining a non-zero fixed point value.  In this case
we obtain the fixed point values
\begin{equation} \label{baryon1}
R''^*={77\over 102}, \, R_b^*={2\over 17}, \, R_t^*={2\over 17},
\end{equation}
which is a physically acceptable fixed point solution.  
We can also try the possibility $R_b^* = 0$, with all other
Yukawa couplings attaining a non-trivial infra-red fixed point 
value. In this case we find
\begin{equation} 
R''^* = {19\over 24}, \, R_t^*={1\over 8}, \, R_{\tau}^*=-{3\over 4},
\label{baryon11}
\end{equation}
which is physically unacceptable.

Next we must 
try fixed points with two of the Yukawa couplings having a zero fixed
point value and others attaining a non-trivial fixed point.  We first 
consider the fixed point with $R_\tau^*=0, \, R''^*=0$.  In this case
we obtain the fixed point values for the other couplings as
\begin{equation} \label{baryon2}
R^*_b=R^*_t={1\over 3},
\end{equation}
rendering this infra-red fixed point as an acceptable one.  
The other possibility
is $R_\tau^*=0, \, R_b^*=0$, which is relevant for the case when 
$\tan\beta$ is small.  In this case, we get the fixed point values
\begin{equation} \label{baryon3}
R''^*={19\over 24}, \, R_t^*={1\over 8},
\end{equation}
which is also a theoretically acceptable fixed point.
We note that the fixed point values in (\ref{baryon1}), (\ref{baryon2})
and (\ref{baryon3}) are not significantly different
from those obtained in~\cite{ap1}, where the $\tau$-Yukawa coupling
evolution was ignored.

Since there are more than one theoretically acceptable IRFPs in this
case, it is necessary to determine, which, if any, of these fixed points
is more likely to be realized in nature.  To this end, we must examine
the stability of each of the fixed point solutions (\ref{baryon1}),
(\ref{baryon2}) and (\ref{baryon3}).

The infra-red stability of a fixed point is determined by the sign of
the quantitites
\begin{equation}\label{lambdaequation}
\lambda_i={1\over b_3}\left[\sum_{j=m+1}^n S_{ij} R^*_j - (r_i+b_3)\right],
\end{equation}
for those couplings which have  a fixed point value  zero, 
$R_i^*=0,\, i=1,2,...,m,$ and by the sign of the eigenvalues of the
matrix
\begin{equation}\label{aijmatrix}
A_{ij}={1\over b_3} R^*_i S_{ij}, \, i=m+1,...,n,
\end{equation}
for those couplings  which have a non-trivial IRFP~\cite{allanach},
where $R_i^*$ is the set of the non-trivial 
fixed point values of the
Yukawa couplings under consideration, and $S_{ij}$ is the matrix
appearing in the corresponding renormalization group
equations (\ref{componentequation})
for the ratios $R_i$.  For stability, we require all the eigenvalues
of the matrix (\ref{aijmatrix}) to have negative real parts
(note that the QCD $\beta$-function $b_3$ is negative).
For the infra-red fixed point (\ref{baryon1}), 
we find from Eq.~(\ref{lambdaequation})
\begin{equation}\label{baryon1stability1}
\lambda_1=-{19\over 17}
\end{equation}
corresponding to $R^*_\tau=0$, and for the eigenvalues of 
the matrix (\ref{aijmatrix})
\begin{equation}\label{baryon1stability2}
\lambda_2=-{10\over 51} \, \, \simeq -0.2, \, \, \,
\lambda_3={-273 - \sqrt{43113}\over 306} \, \, \simeq -1.6, \, \, \,
\lambda_4={-273+\sqrt{43113}\over 306} \, \, \simeq -0.2,
\end{equation}
corresponding to the non-trivial fixed point values for $R''^*,\,
R^*_b$ and $R^*_t,$ respectively.  Since all the $\lambda_i,\, i=1,2,3,4$
are negative, the fixed point (\ref{baryon1}) is infra-red stable.

Next we consider the stability of the fixed point (\ref{baryon2}).
Since in this case $R_\tau^*=R''^*=0$, we have to obtain the behaviour
of these  couplings around the origin.  This behaviour is determined by the
quantities (\ref{lambdaequation}) which, in this case,  are
\begin{equation}
\lambda_1=-{4\over 3}, \, \lambda_2={11\over 9},
\end{equation}
thereby indicating that the fixed point is unstable in the infra-red region.
For completeness we also obtain the behaviour of $R_b$ and $R_t$ around
their respective fixed points governed by the corresponding
eigenvalues of the matrix (\ref{aijmatrix}). We obtain for
the eigenvalues of this matrix
\begin{equation}
\lambda_3=-{5\over 9}, \, \lambda_4=-{7\over 9}.
\end{equation}
Although $\lambda_3,\, \lambda_4$ are negative, the fact 
that $\lambda_2>0$ implies that the fixed point
(\ref{baryon2}) is unstable in the infra-red region.
Thus, this infra-red fixed point with trivial
fixed point value for the baryon number
violating coupling $\lambda_{233}''$ will 
never be realized at low energies, 
and must be rejected.  
This is  to be contrasted with the corresponding result obtained 
in the absence of $B$-violation~\cite{schrempp2}, wherein 
the fixed point (\ref{baryon2}) is infra-red stable. Thus, the
inclusion of the baryon number violating coupling
has the effect of making the fixed point (\ref{baryon2}) unstable. 

Similarly,  it is straightforward to see that the fixed point
(\ref{baryon3}) is  unstable in the infra-red, and must be rejected.
We note that this  is in contrast to the result found in
~\cite{allanach2} where the fixed point (\ref{baryon3})  was found to
be stable. The reason for this difference lies in the
fact that in ~\cite{allanach2} the bottom- and the
$\tau$-Yukawa couplings,  $h_b$ and $h_{\tau}$,
were completely ignored.

At this stage it is interesting to ask whether the 
baryon number violating
stable fixed point solution (\ref{baryon1}) would still be  stable
if there were lepton number violation, but with trivial infra-red
fixed point value for the lepton number violating coupling. We first 
consider the case with the lepton number violating
coupling $\lambda_{233}$ being present  with 
$R^*=R_\tau^*=0$, and the  rest of the  couplings approaching the 
fixed point values (\ref{baryon1}). Stability analysis then yields
for the eigenvalues (\ref{lambdaequation})
\begin{equation}
\lambda_1 = -1, \, \lambda_2 = -{19 \over 17}
\label{referee1}
\end{equation}
corresponding to $R^*=R_\tau^*=0$, and for the eigenvalues of 
(\ref{aijmatrix}) corresponding to the nontrivial fixed
point (\ref{baryon1}) the results are as in (\ref{baryon1stability2}). 
This shows that the  baryon number violating fixed point
(\ref{baryon1}) remains stable in presence of  trivial
fixed point value for the lepton number violating coupling
$\lambda_{233}$.

We next consider the 
baryon number violating
stable fixed point solution (\ref{baryon1}) with lepton number
violating coupling $\lambda'_{333}$, but with this lepton
number violating coupling approaching a trivial fixed point,
i.e. $R'^* = R_{\tau}^* = 0$, and the other Yukawa couplings 
attaining the fixed point values (\ref{baryon1}). In this case
stability analysis leads for the eigenvalues (\ref{lambdaequation})
\begin{equation}
\lambda_1 = 0, \, \lambda_2 = -{65 \over 51}
\label{referee2}
\end{equation}
corresponding to $R'^*=R_\tau^*=0$, and for the eigenvalues of 
(\ref{aijmatrix}) corresponding to the nontrivial fixed
point (\ref{baryon1}) the results are as in (\ref{baryon1stability2}). 
We conclude that such a fixed point is a saddle point or ultra-violet 
fixed point, and will never be realized in the infra-red region.
Thus, the fixed point (\ref{baryon1}) ceases to be a stable
infra-red stable fixed point in the presence of trivial
fixed point value for the lepton-number
violating coupling $\lambda'_{333}$. The same conclusions are reached
when we have both the lepton number violating couplings,
$\lambda_{233}$ and $\lambda'_{333}$, but with both approaching a
trivial fixed point, and rest of the couplings approaching the
fixed point (\ref{baryon1}).

One may also consider the case where the couplings $\lambda''^*_{233}$,
$h_\tau,\, h_b$ attain trivial fixed point values, whereas $h_t$ attains
a non-trivial fixed point value.  In this case we find $R_t^*=7/18$,
the well-known Pendleton-Ross fixed point~\cite{pendross}.   The stability
of this fixed point solution is obtained by simply considering
\begin{equation}
\lambda_i={1\over b_3} \left[ (S_B)_{i4} R_4^* -(r_i+b_3)\right],\, \,
i=1,2,3,
\end{equation}
which yields
\begin{equation}
\lambda_1=-1, \, \lambda_2={38\over 27}, \, \lambda_3={35\over 54},
\end{equation}
thereby rendering this fixed point unstable.

Finally, one may consider the case where $R''^*=0$, with $R_\tau,\, R_b,
\, R_t$ attaining non-trivial fixed point values.  This is the case
of the MSSM with all the third generation Yukawa couplings taken into
account.  In this case, we find the fixed point solution:
\begin{equation}
R_\tau^*=-{70\over 61}, \, R^*_b={97\over 183}, \, R^*_t={55\over 183},
\end{equation}
which must be rejected as being unphysical.

Thus, we have shown that the only fixed point
which is stable in the infra red region is the
baryon number,  and $R_p$,  violating solution (\ref{baryon1}).  
We note that the value
of $R_t^*$ corresponding to this solution is lower than the Pendleton-Ross
fixed point value of $7/18$ for MSSM with baryon number,
and $R_p$,  conservation. 

\bigskip

\subsubsection{Fixed Points with $L$ violation}

Having established a stable infra-red fixed point with baryon number
violation, we now investigate the possibility of having stable fixed
points with lepton number violation.  We first consider the possibility
of  having simultaneous fixed points for $h_\tau,\, h_b,\, h_t$,
and the two lepton number violating couplings $\lambda_{233}$ and
$\lambda'_{333}$.  Ordering the ratios of the squares of
the Yukawa coupings to the square of the gauge coupling $g_3$ as
$R_i=(R,R',R_\tau,R_b,R_t)$,  the anomalous dimension matrix,
 $S_L$, for this case is given by  
\begin{equation}
S_L=\left[
\begin{array}{c c c c c}
4 & 3 & 4 & 0 & 0 \\
1 & 6 & 1 & 6 & 1 \\
4 & 3 & 4 & 3 & 0 \\
0 & 6 & 1 & 6 & 1 \\
0 & 1 & 0 & 1 & 1 \\
\end{array}
\right],
\label{lepton1}
\end{equation}
leading to the fixed point values
\begin{equation}
R^*=0, \, R'^*={97\over 183}, \, R^*_\tau=-{70\over 61}, \, R^*_b=0,
\, R^*_t={55\over 183}.
\end{equation}
Since $R^*_\tau<0$, this fixed point is unacceptable.
Thus, a simultaneous IRFP for both the $L$ violating couplings
$\lambda_{233},\, \lambda'_{333}$, and the third generation Yukawa
couplings is not possible.

We now consider the possibility of the two $L$ violating couplings
separately, i.e. we shall take either $\lambda_{233}\ll \lambda'_{333}$,
or $\lambda'_{333}\ll \lambda_{233}$, respectively.  In the first case
we reorder the couplings as $R_i=(R', R_\tau, R_b, R_t)$, so that
the anomalous dimension matrix, denoted as  $S_{L1}$,  now reads
\begin{equation}
S_{L1}=\left[
\begin{array}{c c c c}
6 & 1 & 1 & 1 \\
3 & 4 & 3 & 0 \\
6 & 1 & 6 & 1 \\
1 & 0 & 1 & 6 \\
\end{array}
\right].
\end{equation}
Since the determinant of this matrix vanishes, there are no fixed points
in this case.  We, thus,  conclude that a simultaneous fixed point for
$\lambda'_{333},\, h_\tau,\, h_b,\,$ and $h_t$ does not exist.  We note
that the vanishing of the determinant corresponds to a solution with
a fixed line or surface.

It must also be noted that even if we were to consider a trivial fixed
point for $R_\tau,\, R^*_\tau=0$, we still have a singular anomalous
dimension matrix.  Thus, there are no fixed points for the $L$ violating
couplings $\lambda'_{333}$, and $h_b$ and $h_t$, with a trivial fixed
point for the $\tau$-Yukawa coupling.  We must, however, consider a 
trivial fixed point for $h_b$, and non-trivial fixed point for other
couplings.  We then obtain the fixed point values
\begin{equation}
R^*_b=0, \, \, R'^*={97\over 183}, \, \, R^*_\tau=-{70\over 61}, \, 
\, R^*_t={55\over 183},
\end{equation}
which must, however, be rejected as unphysical.

We must now consider a situation where two of the couplings
approach a trivial fixed point value, with the other two approaching
a non-trivial fixed point value.
We first consider trivial fixed points for 
$h_\tau$ and  $h_b$, with non-trivial
fixed points for $\lambda'_{233}$ and $h_t$,
which is relevant for low values of
$\tan\beta$.  In this case we obtain the
fixed point values
\begin{equation}\label{lambdaprimefinal}
R^*_b=R_\tau^*=0, \, R'^*=R^*_t={1\over 3}. 
\end{equation}
We must now study the stability of the fixed point solution 
(\ref{lambdaprimefinal}).  The stability of $R^*_b = R^*_\tau=0$ is
determined by the sign of the quantities $\lambda_i$
in (\ref{lambdaequation}), which are calculated to be
\begin{equation}
\lambda_1=0,\, \lambda_2=-{4\over 3},
\end{equation}
from which we conclude that the fixed point (\ref{lambdaprimefinal})
will never be reached in the infra-red region.  The fixed point is
either a saddle point or an ultra-violet fixed point.  
Next we consider the case in which $\lambda'_{333}$ and
$h_\tau$  approach  trivial fixed point values, whereas
$h_b$ and $h_t$ attain nontrivial  fixed points.  We obtain
\begin{equation}\label{lambdaprimefinal2}
R'^*=R_\tau^*=0, \, \,  R_b'^*=R^*_t={1\over 3}.
\end{equation}
The stability of the fixed point  (\ref{lambdaprimefinal2}) is
determined in a manner annalogous to that of the  fixed point
(\ref{lambdaprimefinal}). We find that the fixed point
(\ref{lambdaprimefinal2}) is either a saddle point, or an
ultra-violet stable fixed point. That the 
stability properties of the fixed points
(\ref{lambdaprimefinal}) and (\ref{lambdaprimefinal2})
are identical is a consequence of the symmetry of
the renormalization group 
equations ~(\ref{rtequation}) - (\ref{rdoubleprimeequation}).
We conclude that there are no non-trivial stable fixed points in the 
infra-red region for the $L$ violating coupling $\lambda'_{333}$.

Finally we consider the case when $\lambda'_{333}\ll \lambda_{233}$.
We reorder the couplings as $R_i=(R,R_\tau,R_b,R_t)$, with the
anomalous dimension matrix $S_{L2}$ given by
\begin{equation}\label{sl2}
S_{L2}=\left[
\begin{array}{c c c c}
4 & 4 & 0 & 0 \\
4 & 4 & 3 & 0 \\
0 & 1 & 6 & 1 \\
0 & 0 & 1 & 6 \\
\end{array}
\right],
\end{equation}
which leads to the fixed point solution
\begin{equation}
R^*=-{97\over 36},  \, R^*_\tau={35\over 18}, \, R^*_b=0,\,
\, R^*_t={7\over 18},
\end{equation}
which must be rejected as a fixed point.
We, therefore, try a fixed point with $R^*_\tau=0$, with the result
\begin{equation}
R^*=-{3\over 4},\, R^*_b={1\over 3}, \, R^*_t={1\over 3},
\end{equation}
which too must be rejected.  We then try a trivial fixed point for
the b-quark Yukawa coupling $R^*_b=0$, and non-trivial fixed points
for all other coupling.  From the matrix (\ref{sl2}), we see that the 
corresponding matrix is singular, so that there are no fixed points.  
We finally try the fixed point with  $R_b^* = R^*_\tau=0$.  For this case
we obtain the solution
\begin{equation}
R_b^* = R^*_\tau=0, \,
R^* = -{3\over 4}, \,  
\, R^*_t={7\over 18},
\end{equation}
which too is unacceptable.  We have checked that the trivial fixed
point for $\lambda_{233},\, h_b, \, h_\tau$, and the Pendleton-Ross
type fixed point for $h_t$,  is unstable in the infra-red region.
We, therefore, conclude that there are no fixed point solutions for
the $L$ violating coupling $\lambda_{233}$.

To sum up, we have found that there are no IRSFPs in the MSSM with
the highest generation lepton number  violation.  
This result, together with the result on the fixed point with 
baryon number violation, shows that
only the simultaneous non-trivial fixed point for 
the baryon number violating coupling
$\lambda''_{233}$, and the top- and bottom-quark Yukawa couplings,
$h_t$ and $h_b$,  is stable in the infra-red region.

It is appropriate to examine the implications of the value  of  the
top-quark mass predicted by our fixed point analysis. From (\ref{baryon1}),
it is readily seen that the fixed point value for the
top-quark Yukawa coupling translates into a top-quark (pole) mass
of about $m_t \simeq 70\sin\beta$ GeV, which is incompatible with the
measured value of~\cite{pdg} of top mass, $m_t \simeq 174$ GeV, for any 
value of  $\tan\beta$. It follows that the true fixed point
obtained here provides only a lower bound on the baryon number
violating coupling $\lambda_{233}'' \stackrel{>}{\sim} 0.97$.

\subsection{Infra-red Fixed Points for the Trilinear Soft Supersymmetry
Breaking Parameters}

We now consider the renormalization group evolution and the fixed point
structure for the  soft supersymmetry breaking trilinear parameters
$A_i$.  As we have seen  previously, there is only one IRSFP in
the MSSM with $B$ and $L$ violation.  We shall, therefore, consider the
IRFPs for the $A$ parameters corresponding to this case only, i.e.
for $A_t$, $A_b$, $A_\tau$ and $A_{\lambda''}$.

Retaining only these parameters, and defining $\tilde{A}_i=A_i/M_3
\, (A_i=A_t, \, A_b, \, A_\tau, \, A_{\lambda''})$, we obtain from
Eq.~(\ref{rge7}) - Eq.~(\ref{rge12}) the relevant renormalization group
equations for $\tilde{A}_i$ (neglecting the $SU(2)_L$ and $U(1)_Y$ 
gauge couplings):
\begin{eqnarray}
& \displaystyle {d \tilde{A}_t\over d(-\ln \mu^2)}=
\tilde{\alpha}_3 \left[ {16\over 3} - (6 R_t -b_3) \tilde{A}_t-
R_b \tilde{A}_b -R'' \tilde{A}_{\lambda''} \right] & \label{atildet}, \\
& \displaystyle {d \tilde{A}_b\over d(-\ln \mu^2)}=
\tilde{\alpha}_3 \left[ {16\over 3} - R_t  \tilde{A}_t-
(6 R_b- b_3) \tilde{A}_b - R_\tau \tilde{A}_\tau -
2R'' \tilde{A}_{\lambda''} \right] & \label{atildeb}, \\
& \displaystyle {d \tilde{A}_\tau\over d(-\ln \mu^2)}=
\tilde{\alpha}_3 \left[ 
-3 R_b \tilde{A}_b - (4 R_\tau -b_3) \tilde{A}_\tau \right] & 
\label{atildetau}, \\
& \displaystyle {d \tilde{A}_{\lambda''}\over d(-\ln \mu^2)}=
\tilde{\alpha}_3 \left[ 8 - 2 R_t  \tilde{A}_t- 2 R_b \tilde{A}_b 
-(6 R''-b_3) \tilde{A}_{\lambda''} \right], & \label{atildelambdadpr} 
\end{eqnarray}
which can be written in the following compact form:
\begin{equation}\label{acomponent}
{d \tilde{A}_i\over d(-\ln \mu^2)}=\tilde{\alpha}_3\left[r_i
-\sum_j K_{ij} \tilde{A}_j \right],
\end{equation}
where $r_i$ have been defined in the discussion  following 
Eq.~(\ref{componentequation}), and where $K$ is a matrix whose
entries are fully specified by the wave function anomalous dimensions
and $R_i$.  A fixed point for $A_i$ is reached when the right hand side
of Eq.~(\ref{acomponent}) vanishes for all $i$.  Denoting this fixed
point solution by $\tilde{A}^*_i$, we have
\begin{equation}\label{afixedpoint}
r_i-\sum_j K^*_{ij} \tilde{A}_j^*=0,
\end{equation} 
where $K^*$ is the matrix $K$ evaluated when $R_i$ take their
fixed point values $R^*_i$.  With the ordering $\tilde{A}_i=
(\tilde{A}_\tau, \tilde{A}_{\lambda''}, \tilde{A}_b, \tilde{A}_t)$,
we see from Eq.~(\ref{atildet}) - Eq.~(\ref{atildelambdadpr}) 
that the matrix $K$  can be written as 
\begin{equation}
K^*=\left[
\begin{array}{c c c c}
4 R_\tau^*-b_3 & 0 & 3 R_b^* & 0 \\
0 & 6 R''^*-b_3 & 2 R_b^* & 2 R_t^*  \\
R_\tau^* & 2 R''^* & 6 R_b^* -b_3 & R_t^* \\
0 & 2 R''^* & R^*_b & 6 R_t^* - b_3 \\
\end{array}
\right],
\end{equation}
with $R_\tau^*=0$, and $R''^*, R^*_b, R^*_t$ given by their fixed point
values in (\ref{baryon1}).  The fixed points $\tilde{A}_i^*$ are given by
the solution of 
\begin{equation}\label{asolution}
\tilde{A}_i^*=\sum_j (K^{*-1})_{ij} r_j,
\end{equation}
with the result
\begin{equation} \label{truetrilinear}
\tilde{A}_\tau^*=-{2\over 17}, \, \, \tilde{A}_{\lambda''}^*
=\tilde{A}_b^*=\tilde{A}_t^*=1.
\end{equation}
We note that the fixed point values for $\tilde{A}_b$ and 
$\tilde{A}_t$ are the same as in MSSM with baryon number 
conservation~\cite{jack}. However, the fixed point
value of the $A$ parameter corresponding to the $\tau$-Yukawa 
coupling is affected by the presence of the $B$-violating parameter
$\tilde{A}_{\lambda''}$.  We further note that the fixed point
(\ref{truetrilinear}) is infra-red stable because of the general
results conneting the stability of a set of $A$ parameters to the 
stability of the corresponding set of Yukawa couplings~\cite{jack}.

\section{Quasi-Fixed Points}

The infra-red fixed points 
that we have discussed in the previous section
are the true IRFPs of the renormalization group equations.
However, these fixed points may not be reached in practice, the
range between the GUT scale and the weak scale being too small
for the ratios to closely approach the fixed point values.  In that case,
the various couplings may be determined by the quasi-fixed point 
behaviour~\cite{hill}, where the value of various couplings at
the weak scale is independent of its value at the GUT scale, provided
the Yukawa couplings at the unification scale are large.  In this
section, we shall discuss the quasi-fixed point behaviour of the
Yukawa couplings and the $A$ parameters of the MSSM with $B$~(and
$R_p$) violation.

Before proceeding with the discussion of quasi-fixed point behaviour
for the MSSM with baryon and lepton  number
(and $R_p$) violation, it is instructive to consider the case 
of MSSM with $R_p$ conservation, and with only one dominant 
Yukawa and gauge coupling, 
$h_t$ and $g_3$, respectively. In this case the renormalization 
group equation (\ref{rge1}) for the top-quark Yukawa coupling can 
be written as
\begin{equation}
{dh_t^2\over d\ln \mu}={h_t^2\over 8 \pi^2}
\left(6 h_t^2 - {16\over 3} g_3^2 \right). \label{rgeapprox}
\end{equation}
If at some large scale $\mu = \Lambda$, the top-quark Yukawa coupling
is large compared to the strong gauge coupling, $h_t^2(\Lambda) 
\gg g_3^2(\Lambda)$, then the running of $h_t^2$ is driven by the
term $6h_t^2$ in the right hand side of (\ref{rgeapprox}).
Thus, we can replace $16g_3^2/3$ in (\ref{rgeapprox})
by some constant average value $16\overline{g_3^2}/3$.
In the evolution of $h_t^2$, with $\mu$ running towards the
infra-red region, a transient slowing down 
in the running of $h_t^2$ is expected in the region
of $\mu$ where the right hand side of (\ref{rgeapprox}),
and consequently $dh_t^2/d\ln \mu$, vanishes, leading
to the quasi-fixed point~\cite{hill} for the
top Yukawa coupling~\cite{carena}
\begin{equation}
h_t^{2*} \approx {{8}\over{9}}\overline g_3^2.
\label{quasi}
\end{equation}
It is clear that the quasi-fixed point (\ref{quasi}) is not a
genuine  fixed point, since its position depends on the large scale 
$\Lambda$ as well as the initial $g_3^2(\Lambda)$, and it is not 
attractive for smaller initial values of $h_t(\Lambda)$. 
In practice, of course, one must evolve $h_t^2(\mu)$, 
together with $g_3^2(\mu)$, for large initial values of $h_t$ to find
the quasi infra-red fixed point. We now apply this procedure to
MSSM with all the third generation Yukawa couplings, and with
baryon and lepton number violation.

\subsection{Yukawa couplings}

Since the simultaneous fixed point for the third generation Yukawa 
couplings and the baryon number violating coupling 
$\tilde{A}_{\lambda''}$ is stable, we shall consider
the quasi-fixed points for these couplings only,
{\it viz.} for the couplings $h_t,\, h_b, h_\tau$ and
$\lambda''_{233}$.  Similarly, we shall consider the 
quasi-fixed points for the corresponding $A$ parameters only.
For this purpose, we define
\begin{eqnarray}
Y_i={h_i^2\over 4\pi}, \, i=t, \, b, \, \tau, \\
Y''={\lambda''^2_{233}\over 4\pi},
\end{eqnarray}
and write the RG equations for these quantities as 
\begin{eqnarray}
& \displaystyle {d Y_t \over d(\ln\mu)}={1\over 2 \pi} Y_t
\left ( 6 Y_t + Y_b+2 Y''-{16\over 3}\alpha_3-3 \alpha_2 -{13\over 15}
\alpha_1 \right),&  \\
& \displaystyle {d Y_b \over d(\ln\mu)}={1\over 2 \pi} Y_b
\left(  Y_t + 6 Y_b+ Y_\tau+2 Y'' %\right. &  \nonumber \\
%& \displaystyle \left.
 -{16\over 3}\alpha_3-3 \alpha_2 -{7\over 15} 
\alpha_1 \right),& \\
& \displaystyle {d Y_\tau \over d(\ln\mu)}={1\over 2 \pi} Y_\tau
\left ( 3 Y_b+4 Y_\tau-3 \alpha_2 -{9\over  5}
\alpha_1 \right),&  \\
& \displaystyle {d Y'' \over d(\ln\mu)}={1\over 2 \pi} Y''
\left ( 2 Y_t +2 Y_b+6 Y''-8\alpha_3 -{4\over  5}
\alpha_1 \right).& 
\end{eqnarray}
The existence of the quasi-fixed point requires~\cite{barger}
\begin{equation}\label{approximatecondition}
{d Y_t \over d(\ln\mu)}\simeq{d Y_b \over d(\ln\mu)}
\simeq{d Y_\tau \over d(\ln\mu)}\simeq{d Y'' \over d(\ln\mu)}\simeq 0.
\end{equation}

The solution of these conditions leads to 
\begin{equation}
Y_\tau^*=-{3\over 730}\left[121 \alpha_1 +100 (\alpha_2-\alpha_3)\right]
\simeq -0.02,
\end{equation}
where we have used  $\alpha_1\simeq 0.017,\, \alpha_2\simeq
0.033,\, \alpha_3\simeq 0.1$ at the effective supersymmetry scale,
which we take to be 1 TeV. This solution is clearly unphysical.  
We are, therefore, led to set $Y_\tau^*=0$, yielding  the solution
\begin{eqnarray}
& \displaystyle Y_t^*={47\alpha_1+225 \alpha_2+200 \alpha_3\over 425}
\simeq 0.066, & \\
& \displaystyle Y_b^*={13\alpha_1+225 \alpha_2+200 \alpha_3\over 425}
\simeq 0.065, & \\
& \displaystyle Y''^*={2\left(11\alpha_1-45 \alpha_2+130 \alpha_3
\right)\over 255}
\simeq 0.092, & 
\end{eqnarray}
which implies 
\begin{eqnarray}
& \displaystyle h_t^*\simeq 0.91, & \label{analyukawatop} \\
& \displaystyle h_b^*\simeq 0.90, & \label{analyukawab}\\
& \displaystyle \lambda^{''*}_{233}\simeq 1.08. & \label{analyukawalambda}
\end{eqnarray}
We note that these quasi-fixed point values  are not significantly 
different from those obtained  in a situation  when $\tau$-Yukawa
coupling is ignored~\cite{barger}.

\subsection{Trilinear Couplings}
We now turn out attention to the renormalization group equations 
(\ref{rge7}) - (\ref{rge12}) for the $A$ parameters, and their
quasi-fixed points.  Since the quasi-fixed point of interest
is for $h_\tau=0$, we cannot determine the quasi-fixed point
value for $A_\tau$ and therefore ignore Eq.~(\ref{rge9}).
In the remaining equations, we substitute $h_\tau=0$, and obtain
the equations that the remaining $A$ parameters must
satisfy in order for us to determine the quasi-fixed point solution:
\begin{eqnarray}
6 Y_t A_t + Y_b A_b + 2 Y'' A_{\lambda''}-{16\over 3}\alpha_3 M_3
-3 \alpha_2 M_2 -{13\over 15} \alpha_1 = 0, & \\
Y_t A_t + 6 Y_b A_b + 2 Y'' A_{\lambda''}-{16\over 3}\alpha_3 M_3
-3 \alpha_2 M_2 -{7\over 15} \alpha_1 = 0, & \\
2 Y_t A_t + 2 Y_b A_b + 6 Y'' A_{\lambda''}-8\alpha_3 M_3
-{4\over 5} \alpha_1 = 0. & 
\end{eqnarray}
These yield the following quasi-fixed point solution:
\begin{eqnarray}
& \displaystyle A_t^*={47\alpha_1 M_1+225 \alpha_2 M_2+200 \alpha_3 M_3
\over 425 Y_t^*}\simeq 0.77 m_{\tilde{g}},& \label{analtrilineartop} \\
& \displaystyle A_b^*={13\alpha_1 M_1+225 \alpha_2 M_2+200 \alpha_3 M_3
\over 425 Y_b^*}\simeq 0.78 m_{\tilde{g}},& \label{analtrilinearb} \\
& \displaystyle A_{\lambda''}^*={2\left(11\alpha_1 M_1-45 \alpha_2 M_2
+130 \alpha_3 M_3\right)\over 255 Y''^*}\simeq 1.02 m_{\tilde{g}}, & 
\label{analtrilinearlambda} 
\end{eqnarray}
where $m_{\tilde{g}}$ is the gluino mass~($ = M_3$) at the weak scale.
We have used the fact that the gaugino masses scale as the
square of the gauge couplings,  and that $\alpha_G\simeq 0.041$ 
at the scale 
$M_G\simeq 10^{16}$GeV.  One must compare these quasi-fixed point 
values with the true fixed-point values (\ref{truetrilinear}).
We note that the quasi-fixed point values (\ref{analtrilineartop}) - 
(\ref{analtrilinearlambda}) provide a lower bound on the
corresponding $A$ parameters.

\section{Numerical Results and Discussion}

In the previous section we have obtained the approximate quasi-fixed point 
values for the Yukawa couplings and the $A$ parameters by an algebraic
solution of the corresponding RG equations.  The RG equations are a set
of coupled first-order  differential equations that must be solved
numerically to obtain accurate values for the fixed points. We have
numerically solved the RG equations for the Yukawa couplings, and the 
$A$ parameters. We now present the results of such a
numerical analysis.

In Fig. 1 we show the fixed point behaviour of the top-quark Yukawa 
coupling as a function of the logarithm of the scale parameter
$\mu$. We have included the evolution equations for the b-quark
as well as the $\tau$-lepton Yukawa couplings, as well as the
B-violating coupling $\lambda_{233}''$, in the numerical
solution. It is seen that for all $h_t \stackrel{>}{\sim} 1$
at the GUT scale, the top-quark Yukawa coupling approaches its
quasi-fixed point at the weak scale. We note that the numerical
evolution of fixed point approaches but does not exactly reproduce
the approximate analytical value in (\ref{analyukawatop}). In Figs. 2 and 3
we present the corresponding approach to the infra-red fixed point for 
the couplings $h_b$ and $\lambda_{233}''$, respectively. These infra-red fixed
points provide a model independent 
theoretical upper bound on the $B$-violating
coupling $\lambda_{233}''$.

In Figs. 4, 5 and 6, we present the fixed point behaviour of the 
corresponding $A$ parameters. We notice the remarkable focussing
property seen in the fixed point behaviour of all the $A$ parameters.
Again, we notice that the numerical evolution of the fixed point
approaches, but does not actually reproduce, the approximate analytical
values of Eqs.~(\ref{analtrilineartop}), (\ref{analtrilinearb}),  and
(\ref{analtrilinearlambda}).  Since the quasi-fixed
point value for the $A$ parameter is inversely
proportional to the quasi-fixed point value of the Yukawa coupling,
it provides a lower bound on the corresponding 
$A$ parameter.

\section{Summary and Conclusions}

We have carried out a detailed renormalization group analysis of the
MSSM with all the third generation Yukawa couplings, and with
highest generation
baryon and lepton number violation. We have shown that the simultaneous
fixed point for the top- and bottom-Yukawa couplings, and the $B$-violating
coupling $\lambda_{233}''$, is the only fixed point that is
stable in the infra-red region. However, the top-quark mass
predicted by this fixed point is incompatible with measured value of the 
top mass. This fixed point, therefore, provides a process-independent
lower bound on the baryon number violating coupling at the 
electroweak scale.

We have shown that all other possible fixed point  solutions are 
either unphysical,  or unstable, in the infra-red region. In particular
there is no infra-red fixed point with simultaneous $B$ and $L$ violation.

We have also carried out the renormalization group analysis of the 
corresponding trilinear soft supersymmetry breaking parameters.
We have obtained the true fixed points for these parameters,
which serve as  upper bounds on these parameters.

Since the true fixed points may not be reached in practice at the 
electroweak scale, we have also obtained the quasi-fixed points of the 
Yukawa couplings and the trilinear parameters. The quasi-fixed point
values for the Yukawa couplings are numerically 
very close to the values obtained
previously by ignoring the $\tau$ Yukawa coupling. 
Since the quasi-fixed points are  reached for large initial
values of the couplings at the GUT scale, these reflect on the
assumption of  perturbative unitarity, or the
absence of Landau poles, of the corresponding couplings.
These quasi-fixed points, therefore,   provide an upper bounds on the 
relevant Yukawa couplings, especially the baryon number violating
coupling $\lambda_{233}''$.  From the true fixed point and the quasi-fixed
point ananlysis we are able to constrain the baryon number violating
coupling $0.97 \stackrel{<}{\sim} \lambda_{233}'' \stackrel{<}{\sim} 
1.08$ in a model independent manner.

We have complemented the quasi-fixed point analysis of the Yukawa couplings 
with an  analysis of the corresponding soft supersymmetry
breaking trilinear couplings. We have shown that the $A$ parameters
for the top- and bottom-quark Yukawa couplings, and the baryon number
violating couplings all show  striking convergence properties.
This strong focussing property is quite independent of
the input parameters at the
unification scale (or equivalently the pattern of supersymmetry breaking),
and the $A$ parameters are, therefore, fully determined
in the quasi-fixed regime. However, the actual values of the $A$
parameters in the quasi-fixed regime are significantly different
from the case when $B$-violating Yukawa couplings are ignored.
In particular, we have constrained  the $A$ parameters to be
$0.77 \stackrel{<}{\sim} A_t/m_g \stackrel{<}{\sim} 1$,
$0.78 \stackrel{<}{\sim} A_b/m_g \stackrel{<}{\sim} 1$,
and $A_{\lambda''}/m_g \simeq 1$.

\medskip

\noindent{\bf Acknowledgements:}  The work of 
PNP is supported by  the University Grants
Commission Research Award. He would like to thank the Inter-University
Centre for Astronomy and Astrophysics, Pune, India for its hospitality
while part of this work was done.

\newpage

\noindent{\bf Figure Captions}

\bigskip

\noindent {\bf Fig. 1.}  Renormalization group evolution of the
top-quark Yukawa coupling $h_t$ as a function of the logarithm
of the energy scale. We have taken the initial values of
$h_t$ at the scale $M_G \sim 10^{16}$ to be $5.0, \, 4.0, \, 3.0, \,
2.0$, and $1.0$.  The initial values of other Yukawa couplings 
are $h_b=0.91,\, h_\tau=0,\,$  and $\lambda''_{233}=1.08$.

\medskip

\noindent {\bf Fig. 2.}  Renormalization group evolution of the
bottom-quark Yukawa coupling $h_b$ as a function of the logarithm
of the energy scale. The initial values 
of $h_b$ at $M_G$ are $5.0, \, 4.0, \, 3.0, \, 2.0,$ and $1.0.$
The initial values of  other Yukawa couplings are 
$h_t=0.92,\, h_\tau=0,\, \lambda''_{233}=1.08$. 

\medskip

\noindent {\bf Fig. 3.}  Renormalization group evolution of the
baryon number violating Yukawa coupling $\lambda''_{233}$ 
as a function of the logarithm of the energy scale.
The initial values are $\lambda''_{233} =  5.0, \,
4.0, \, 3.0, \, 2.0, \, 1.0$.
The initial values of other Yukawa couplings 
are $h_t=0.92,\, h_b=0.91,\, h_\tau=0$.

\medskip

\noindent {\bf Fig. 4.}  Renormalization group evolution of ratio 
$A_t/m_g$ as a function of the logarithm of the energy scale for 
several different initial values at $M_G$.
The initial values for other parameters at $M_G$ are
$h_t=5.0, \, h_b=0.91,\, h_\tau=0,\, \lambda''_{233}=1.08$,
and $A_b/m_g = 1.94, \, A_{\lambda''}/m_g = 2.57$.

\medskip

\noindent {\bf Fig. 5.}  Renormalization group evolution of the
ratio $A_b/m_g$ as a function 
of the logarithm of the energy scale.
Other parameters at the scale $M_G$ are
$h_b=5.0, \, h_t=0.92,\, h_\tau=0,\, \lambda''_{233}=1.08$,
and $A_t/m_g=1.92, \, A_{\lambda''}/m_g=2.57$.

\medskip

\noindent {\bf Fig. 6.}  Renormalization group evolution of the
trilinear coupling $A_{\lambda''}/m_g$.
Other parameters at the scale $M_G$ are
$\lambda''_{233}=5.0, \, h_t=0.92,\, h_b=0.92, \,h_\tau=0$,
and $A_t/m_g = 1.92, \, A_b/m_g = 1.94$.

\medskip

\newpage

\begin{center}
\begin{figure}\nonumber
\epsfig{figure=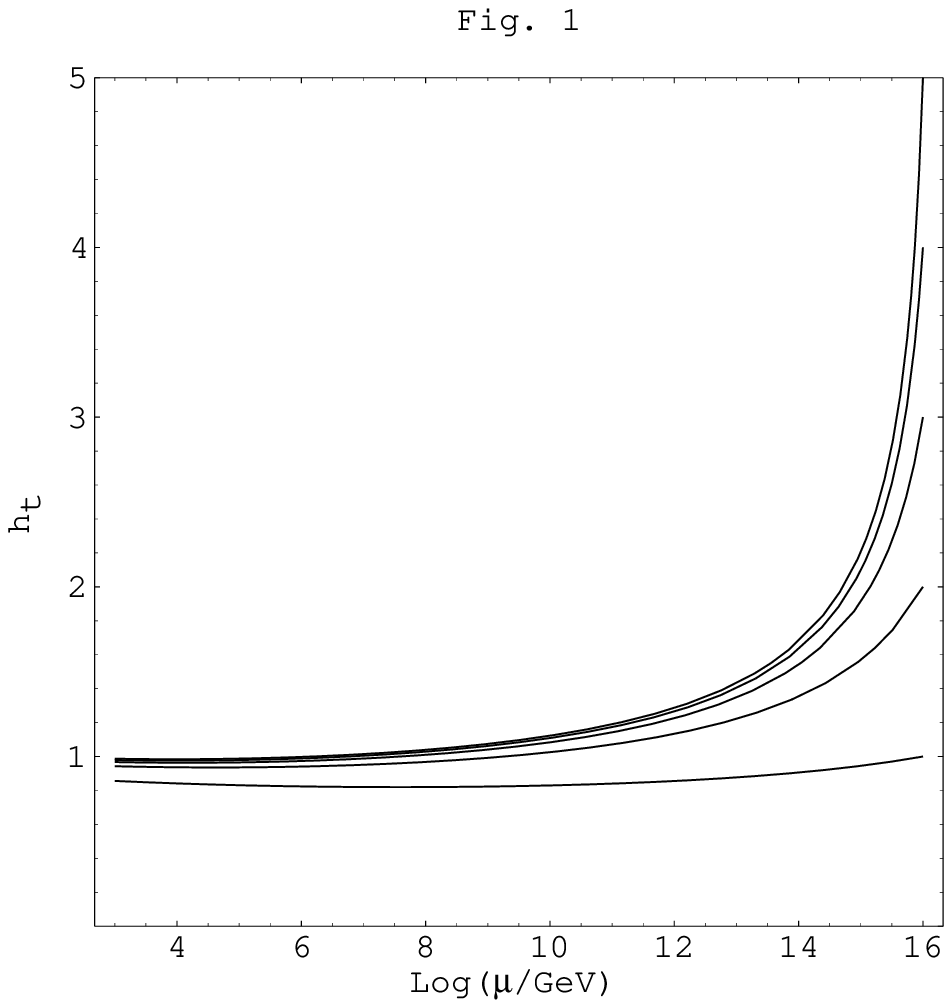,width=10.5cm,height=10.5cm}
\end{figure}

\begin{figure}\nonumber
\epsfig{figure=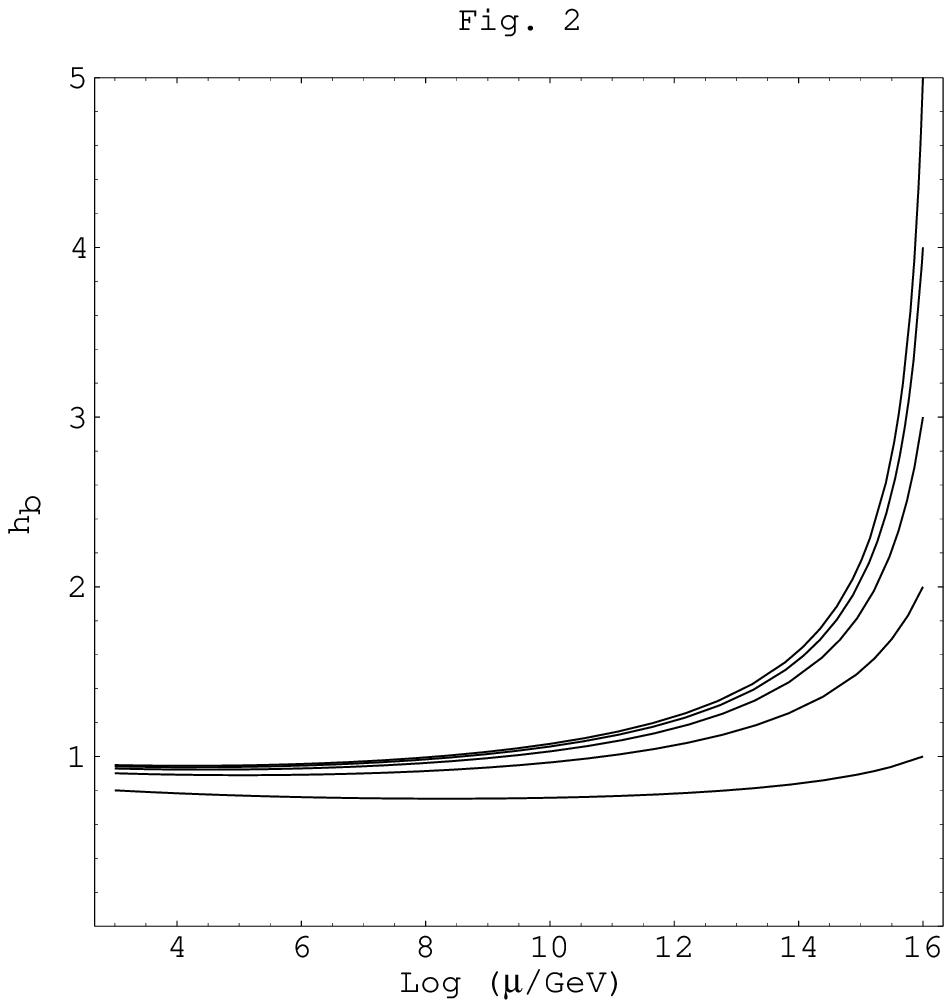,width=10.5cm,height=10.5cm}
%\caption{}
\end{figure}

\newpage

\begin{figure}\nonumber
\epsfig{figure=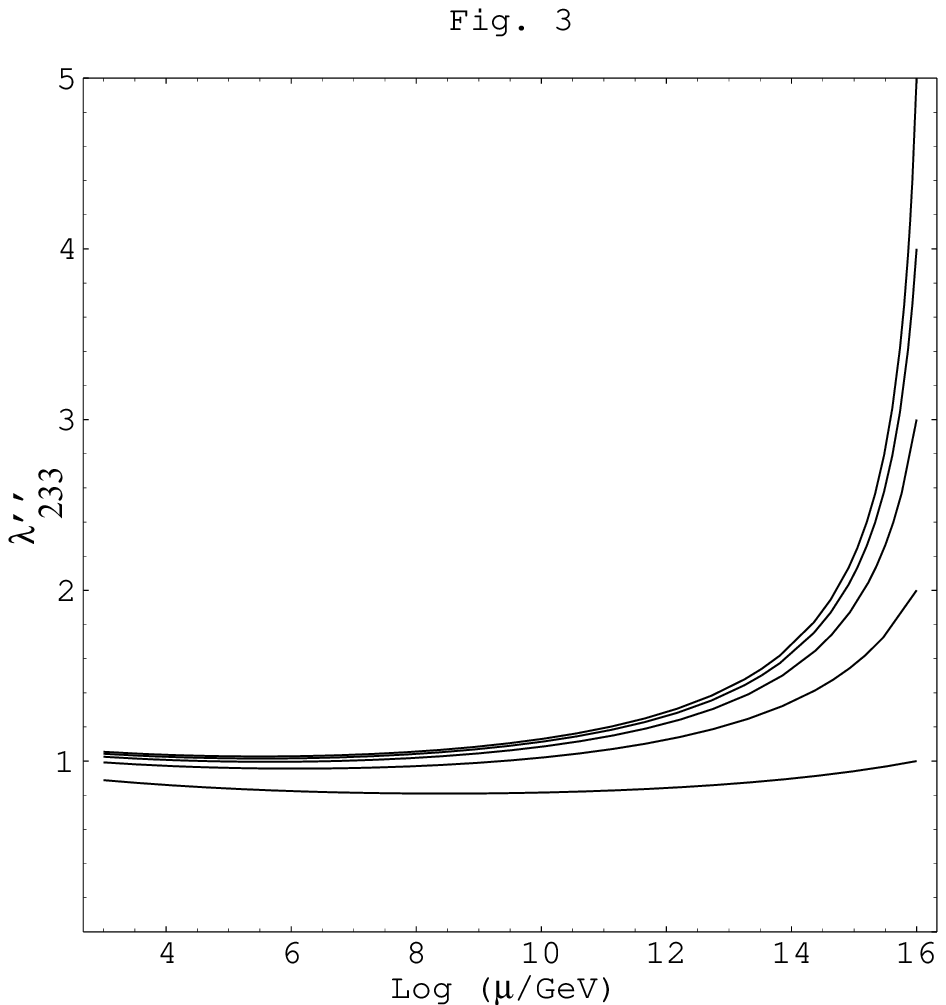,width=10.5cm,height=10.5cm}
\end{figure}

\begin{figure}\nonumber
\epsfig{figure=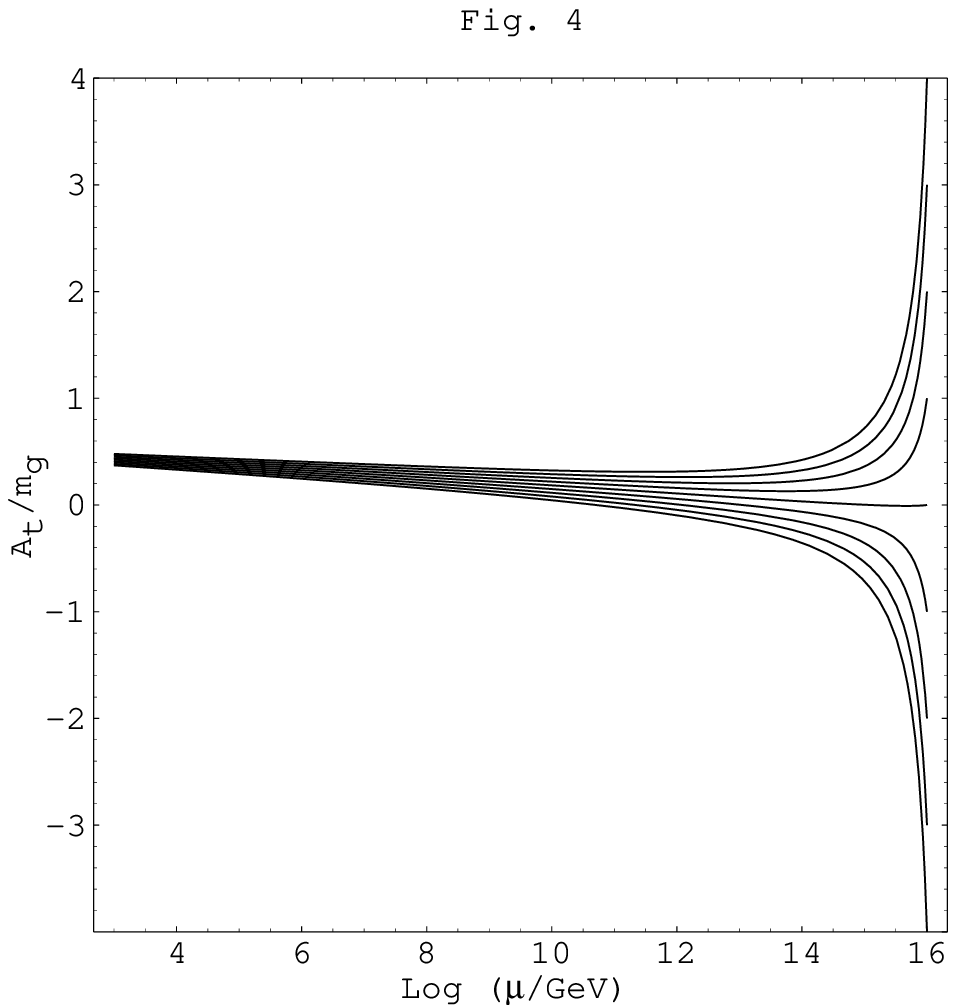,width=10.5cm,height=10.5cm}
\end{figure}

\begin{figure}\nonumber
\epsfig{figure=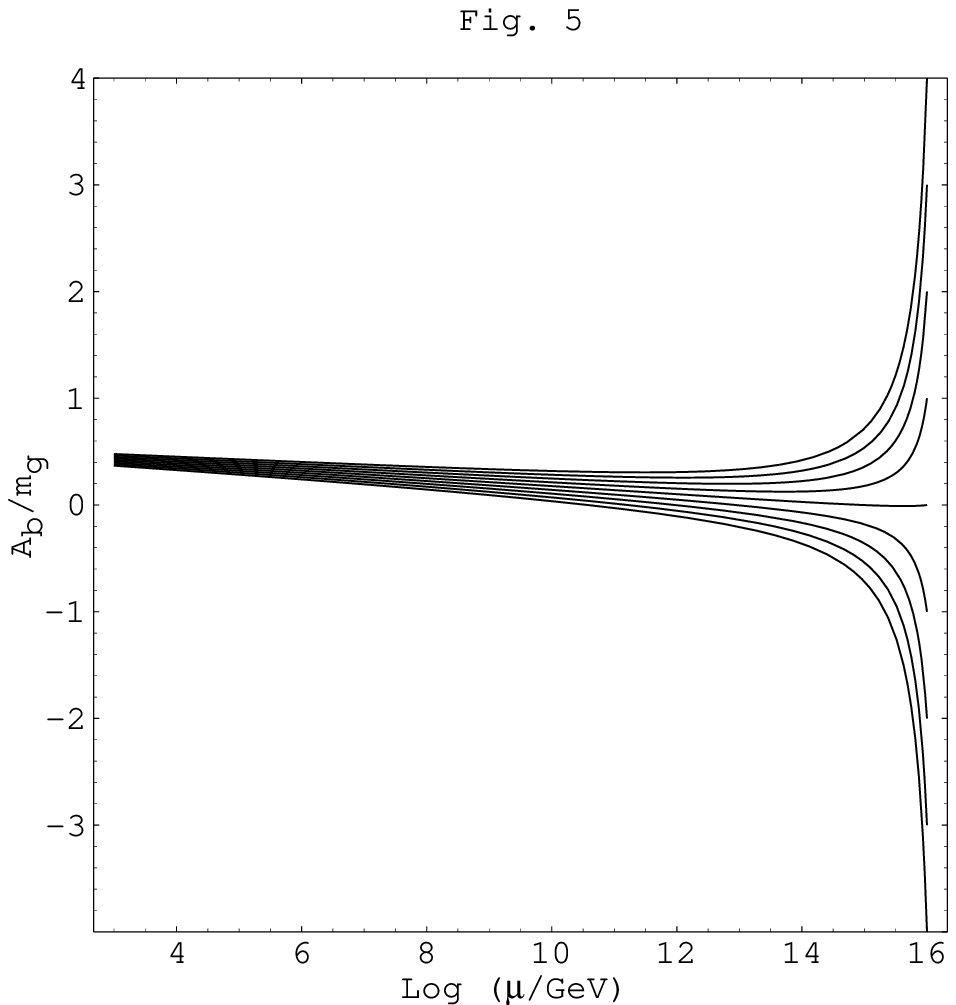,width=10.5cm,height=10.5cm}
%\caption{}
\end{figure}

\begin{figure}\nonumber
\epsfig{figure=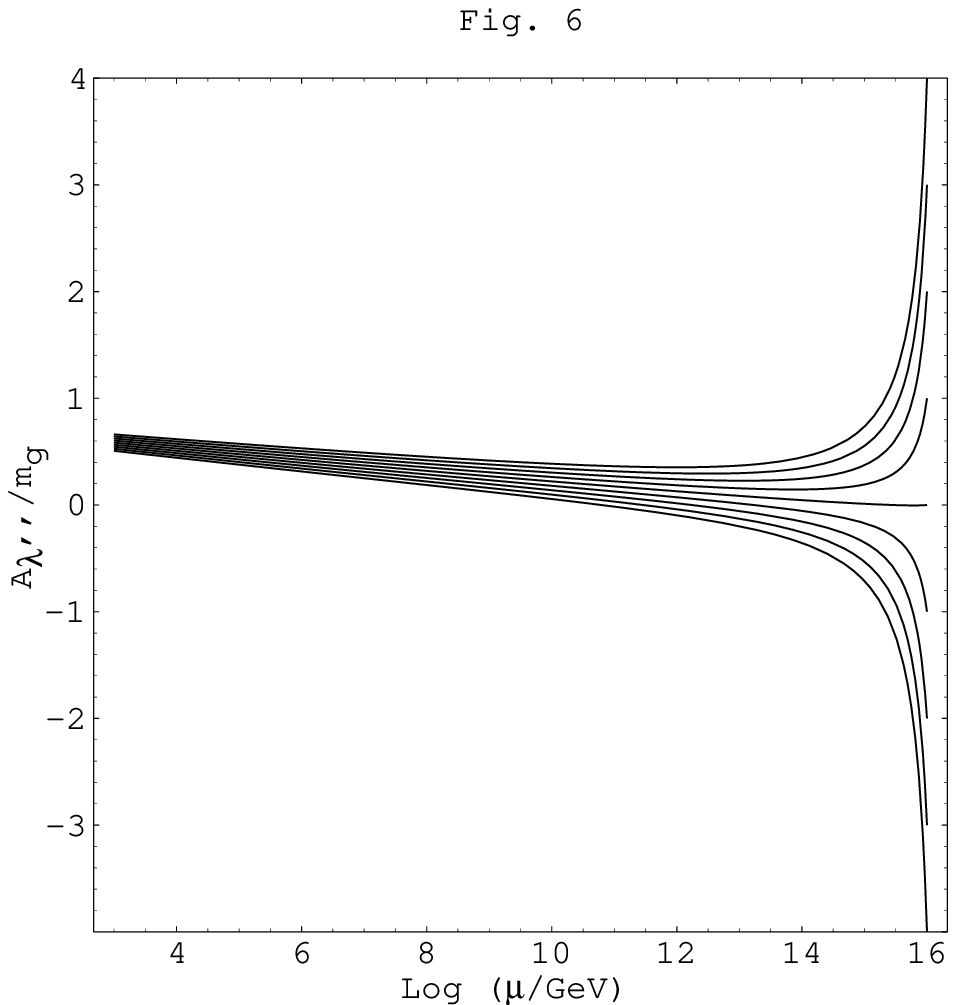,width=10.5cm,height=10.5cm}
\end{figure}

\end{center}

\end{document}